\documentclass[aps,10pt,twocolumn,showpacs,pra,citeautoscript,amsmath,amssymb,floatfix,superscriptaddress]{revtex4-2}
\usepackage{hyperref,amsthm,mathtools,braket,multirow}
\hypersetup{colorlinks=true,linkcolor=blue,citecolor=blue}

\DeclareMathOperator{\Tr}{Tr}

\newcommand{\ketbra}[2]{\ensuremath{\left|#1\right\rangle\!\left\langle#2\right|}}

\newcommand{\tr}[2]{\Tr_{#1}\left( #2 \right)}
\DeclareSymbolFont{bbold}{U}{bbold}{m}{n}
\DeclareSymbolFontAlphabet{\mathbbold}{bbold}
\newcommand{\iden}{\mathbbold{1}}

\theoremstyle{plain}
\newtheorem{thm}{Theorem}
\newtheorem{lem}[thm]{Lemma}

\theoremstyle{definition}

\theoremstyle{remark}

\begin{document}
	
\title{Certifying quantum signatures in thermodynamics and metrology via contextuality of quantum linear response}
\author{Matteo Lostaglio}
\affiliation{ICFO-Institut de Ciencies Fotoniques, The Barcelona Institute of Science and Technology, Castelldefels (Barcelona), 08860, Spain}
\affiliation{QuTech, Delft University of Technology, P.O. Box 5046, 2600 GA Delft, The Netherlands}
\email{lostaglio@protonmail.com}

\begin{abstract}
 We identify a fundamental difference between classical and quantum dynamics in the linear response regime by showing that the latter is in general contextual.   This allows us to provide an example of a quantum engine whose favorable power output scaling \emph{unavoidably} requires nonclassical effects in the form of contextuality.  Furthermore, we describe contextual advantages for local metrology. Given the ubiquity of linear response theory, we anticipate that these tools will allow one to certify the nonclassicality of a wide array of quantum phenomena.
\end{abstract}

\maketitle

Linear response theory describes the reaction of a quantum system to a small perturbation. The theory finds countless applications in many fields of quantum physics, including molecular, atomic and nuclear physics, quantum optics and statistical mechanics. In this paper we present a general test to certify whether the linear response of a quantum system  necessarily requires contextuality.

Using these results, we identify quantum signatures in the power of a heat engine. In the context of quantum thermodynamics, the issue of identifying truly quantum signatures has been a long-standing problem in the field. Several theoretical claims have been made that quantum coherence can offer improvements over certain incoherent thermodynamic engines and refrigerators (e.g., \cite{scully2011quantum, killoran2015enhancing,brask2015small, mitchison2015coherence,uzdin2015equivalence, hardal2017quantum, dorfman2018efficiency, levy2018quantum, camati2019coherence, dann2020quantum} and references therein), followed by recent experimental effort \cite{klatzow2019experimental}.  However, similar signatures can be observed in classical engines as well~\cite{nimmrichter2017quantum, onam2019classical}. Hence, such claims should be backed by a no-go theorem that 1. Defines
a precise notion of nonclassicality 2. Shows that this notion leads to statistical predictions incompatible with the corresponding quantum statistics. 

Here we adopt a stringent notion of nonclassicality, namely generalised contextuality \cite{spekkens2005contextuality}. We prove that the power output of a two-stroke quantum engine in the weak coupling regime cannot be achieved in any noncontextual model reproducing the operational features of the quantum experiment \footnote{\emph{Operational} here refers to statements that can be in principle verified from the  conditional probabilities of observing certain outcomes for given input measurements, without relying on a detailed model of the inner workings of the devices at hand. The terminology refers to the ``black-box'' approach successfully applied in quantum foundations \cite{spekkens2005contextuality}, device-independent quantum cryptography and self-testing~\cite{Supic2020selftestingof}
	}. As a second application, we turn to local metrology and consider the archetypal example of phase estimation using a qubit system. We show that, given the phenomenology of the phase estimation experiment, a nonzero Fisher information is incompatible with all classical (noncontextual) models. This complements a recent result showing that certain features of post-selected metrology are nonclassical \cite{arvidssonshukur2019quantum}, but in our case we do not have to consider any post-selection. 

The tools developed here, applicable as they are to any quantum system in the linear response regime, can find applications in the identification of genuine quantum signatures in a wide range of different platforms. 
  
\emph{Non-contextual ontological models.} Great care needs to be taken when claiming that the performance of a device \emph{requires} nonclassicality. For example, Ref.~\cite{nimmrichter2017quantum} shows that a short-time cooling enhancement, which in quantum theory is attributed to the presence of quantum coherence, also occurs in classical models where a set of oscillators undergoes Hamiltonian evolution.
Here we want to identify signatures in the dynamics of a quantum system which \emph{unavoidably} signal quantum effects are at play. Formally, we will identify phenomena which cannot occur within any noncontextual ontological model (OM). Let us describe in detail this broad class of models~\cite{leifer2014is}.

We may start from the operational description of preparations, transformations and measurements, understood as sets of laboratory instructions according to which these operations are performed. To each, we associate the corresponding physical description in the OM, as summarized in the table below:
\begin{enumerate}
	\item To every preparation procedure $P$ one assigns a probability distribution $\mu_P(\lambda)$ over some (measurable) set of physical states $\lambda$. For example, if $P$ involves leaving the system alone for a long time and $\lambda=(x_1, \dots x_N, p_1, \dots p_N)$ are phase space points, $\mu_P(\lambda)$  may be a thermal distribution. 
	\item A transformation procedure $T$ is described by an update rule giving the probability that any final state $\lambda'$ is reached, given that the initial state was $\lambda$. We denote this transition probability by $\mathcal{T}_T(\lambda'|\lambda)$. For example, $\mathcal{T}_T(\lambda'|\lambda)$ may be generated by a rate equation among a discrete set of $\lambda$'s, as in the classical model described in Ref.~\cite{onam2019classical}. Or, in Hamiltonian dynamics,  $\lambda=(x,p)$ and after time $t$ $\mathcal{T}_T(x',p'|x(0),p(0))= \delta(x' - x(t))\delta(p' - p(t))$, where $(x(t),p(t))$ is the solution of Hamilton's equations with initial conditions $(x(0),p(0))$.
	\item A measurement procedure $M$ with outcomes $k$  is associated to a response function $\xi_M(k|\lambda')$, giving the probability that an outcome $k$ is returned by $M$ if the physical state is $\lambda'$. For example, in classical mechanics, if $M$ is a measurement of the energy $E$ of a single particle of mass $m$ and momentum $p$ in a potential $V(x)$, $\xi_M(E|x,p) = \delta\left(E-\left[\frac{p^2}{2m} + V(x)\right]\right)$. In a general  OM,  $\xi_M(k|\lambda')$ may be nondeterministic.
\end{enumerate}

 Let $p(k|T(P),M)$ be the statistics collected in an experiment where $P$ is prepared, a transformation $T$  is applied and finally a measurement $M$ with outcomes $k$ is performed. The OM predicts
 
\begin{equation}
\label{eq:OMprediction}
p(k|T(P),M) =  \int d\lambda d\lambda' \mu_P(\lambda) \mathcal{T}_T(\lambda'|\lambda)\xi_M(k|\lambda'),
\end{equation}
as naturally follows from the propagation of probabilities. Hamiltonian mechanics is just a member of a class of OM.

Since arbitrary operational statistics, quantum or otherwise, can be reproduced by an appropriate OM, here we consider the subclass of OM that are noncontextual~\footnote{Another property that is typically required is locality, but since here we are not dealing with spacelike separated systems such assumption is inconsequential.}. An OM is noncontextual, in the generalized sense introduced by Spekkens \cite{spekkens2005contextuality, spekkens2019ontological}, when it has the property of \emph{assigning identical physical descriptions to operationally indistinguishable procedures}. Specifically, two preparations $P$, $P'$ are operationally indistinguishable (denoted $P \simeq_{op} P'$) when $p(k|P,M) \equiv p(k|P',M)$ for every measurement procedure $M$. This means no experiment is able to distinguish between $P$ and $P'$. A noncontextual model requires 
\begin{equation}
\label{eq:preparationNC}
P \simeq_{op} P' \Rightarrow \mu_P(\lambda) =  \mu_{P'}(\lambda) \quad \forall \lambda.
\end{equation}
The same has to hold for operationally equivalent measurements and transformations: if we define \mbox{$M \simeq_{op} M'$} as $p(k|P,M) \equiv p(k|P,M') \; \forall P$ and $T \simeq_{op} T'$ as $p(k|T(P),M) \equiv p(k|T'(P),M) \; \forall P, M$, a noncontextual OM is one for which
\begin{align}
M  & \simeq_{op} M'   & \Rightarrow & \quad \xi_M(k|\lambda) =  \xi_{M'}(k|\lambda) \quad \; \forall k, \lambda, \label{eq:measurementNC} \\
T & \simeq_{op} T'  &  \Rightarrow & \quad \mathcal{T}_T(\lambda'|\lambda) =  \mathcal{T}_{T'}(\lambda'|\lambda) \quad  \forall \lambda,\lambda' . \label{eq:transformationNC}
\end{align}
Noncontextuality, in the general form presented here, can be understood as an extension of the original Kochen-Speckers notion (\cite{leifer2014is}, Appendix C). One can easily see that classical Hamiltonian dynamics is a class of noncontextual OM (see Supplemental Material~\ref{app:hamilton} (SM~\ref{app:hamilton})). Noncontextual models include as special cases the classical models previously considered in the literature: e.g., discrete models with jump probabilities generated by rate equations \cite{onam2019classical}; Hamiltonian dynamics obtained via classical limit \cite{nimmrichter2017quantum}; quantum mechanics in a fixed basis obtained via dephasing in the energy basis~\cite{uzdin2015equivalence}. Other examples include Spekken's toy model \cite{spekkens2007evidence} or Hamiltonian mechanics with an uncertainty principle  (the latter  is equivalent to Gaussian quantum mechanics \cite{bartlett2012reconstruction}). These examples show that noncontextual OM allow to reproduce features normally attributed to quantum measurement disturbance, superposition and entanglement.

	\begin{table}[t]
	\begin{tabular}{|c|c|}
		\hline
		Operational description & OM description \\
		\hline \hline
		P  & $\mu_P(\lambda)$ \\
		preparation procedure & probability distribution
		\\
		\hline \hline
		T  & $\mathcal{T}_T(\lambda'|\lambda)$ \\
		transformation procedure & transition probabilities
		\\
		\hline \hline
		M (with outcomes $k$)  & $\xi_M(k|\lambda')$ \\
		measurement procedure & response function
		\\
		\hline \hline
		$p(k|T(P),M)$  & \multirow{2}{*}{  $ \int d\lambda d\lambda' \mu_P(\lambda) \mathcal{T}_T(\lambda'|\lambda)\xi_M(k|\lambda')$} \\
		operational statistics & 
		\\
		\hline
	\end{tabular}
\end{table}

In this paper we will hence adopt the same stringent notion of \emph{quantum signature} used to analyse several quantum information primitives \cite{schmid2018contextual, saha2019state, arvidssonshukur2019quantum,lostaglio2019contextual, tavakoli2020measurement}: a set of \emph{operational features} that unavoidably require contextuality.

\emph{Quantum linear response.} Consider a quantum state $\ket{\psi(t)}$ in a finite-dimensional Hilbert space evolving according to the Schr\"odinger equation under a time-dependent perturbation $V(t)$:
\begin{equation}
\label{eq:unitarydriving}
i \hbar \frac{d}{dt}\ket{\psi(t)} = [H_0 + g V(t)] \ket{\psi(t)}. 
\end{equation}
We develop our considerations here for pure states, but the extension to mixed states by linearity is straightforward. We are interested in the change of expectation value of an observable $O= \sum_k o_k \ketbra{o_k}{o_k}$ due to the perturbation (for technical convenience, without loss of generality we shift $O$ so that $o_k \geq 0$). It is convenient to work in the interaction picture  (label ``I''), $O_I(t) = e^{i H_0 t/\hbar} O e^{-i H_0 t/\hbar}$, $\ket{\psi_I(t)} = e^{i H_0 t/\hbar} \ket{\psi(t)}:= U_I(t)\ket{\psi(0)}$ and study 
\small
\begin{align}
\langle \Delta O_I \rangle^Q_t & :=  \bra{\psi_I(t)} O_I(t) \ket{\psi_I(t)} -  \bra{\psi_I(0)} O_I(t) \ket{\psi_I(0)}. \label{eq: deltaO}
\end{align}
\normalsize
 where ``Q'' stands for ``Quantum''.
Operationally this corresponds to
\small 
\begin{align}
\langle \Delta O_I \rangle_t := \sum_k o_k p(k|T_{t}(P),M_{t})  -  \sum_k o_k p(k|P,M_{t}). 
\label{eq:operationalDeltaO}
\end{align}
\normalsize
where $P$, $T_{t}$ and $M_{t}$ are the preparation, transformation and measurement procedures described in quantum mechanics by  initial state $\ket{\psi(0)}$, unitary dynamics $U_I(t)$ and measurement of the observable $O_I(t)$, respectively. From Dyson's series
\begin{equation}
\label{eq:dyson}
U_I(t) = \iden - \frac{i g}{\hbar} \int_{0}^t dt' V_I(t') + \mathcal{O}(g^2),
\end{equation}
where $V_I(t) =  e^{i H_0 t/\hbar} V(t) e^{-i H_0 t/\hbar}$. Quantum linear response gives
\small
\begin{equation}
\label{eq:linearresponse}
\langle \Delta O \rangle^Q_t = \frac{i g }{\hbar} \int_0^t dt' \bra{\psi(0)}[V_I(t'),O_I(t)]\ket{\psi(0)} + \mathcal{O}(g^2). 
\end{equation}
\normalsize
The most important aspect of this formula is that one can have a response of $\mathcal{O}(g)$ if there are no  pairwise commutations among $\ketbra{\psi(0)}{\psi(0)}$, $O_I(t)$ and $\int dt' V_I(t')$. 

Another crucial fact is encoded in the following channel equality \footnote{A channel is a completely positive and trace preserving map, describing the most general quantum evolution of a system.}. Suppose that for $g$ small enough
\begin{equation}
\label{eq:operatoridentity}
\frac{1}{2} \mathcal{U}_t + \frac{1}{2} \mathcal{U}^\dag_{t} = (1-\tilde{p}_d) \mathcal{I} + \tilde{p}_d \mathcal{C}_t,  
\end{equation}
where $\mathcal{U}_t(\cdot) := U_I(t)(\cdot)U_I^\dag(t)$, $\mathcal{I}$ is the identity channel, $\mathcal{C}_t$ is some other channel and $\tilde{p}_d = \mathcal{O}(g^2)$ as $g \rightarrow 0$. We will later give tools to verify if a quantum experiment under consideration admits this decomposition in linear response. For now it suffices to say that in the case of a single qubit this decomposition holds for every nontrivial perturbation.

Eq.~\eqref{eq:operatoridentity} underlies the fact that the transformation $T_{t}$ (represented by $U_t$ in quantum mechanics) can be reversed, to first order in $g$, by convex combination with another transformation $T^*_{t}$ (represented by $U^\dag_I(t)$ in quantum mechanics). In particular, tossing a fair coin and performing either $T_{t}$ or $T^*_{_t}$ is operationally indistinguishable from doing nothing with probability $1- \mathcal{O}(g^2)$.
These facts can be summarised as:
\begin{equation}
\label{eq:operationalequivalence}
\frac{1}{2} T_{t} +  \frac{1}{2} T^*_{t} \simeq_{op} (1-p_d) T_{\rm id} + p_d T'_{t},
\end{equation}
where $T_{\rm id}$ denotes the `do nothing' operation and $T'_{t}$ denotes some other transformation. As we will see, this approximate `reversibility by mixing' or `stochastic reversibility' tells us that the perturbation $T_{t}$ cannot be `too far' from the do-nothing operation in any noncontextual model.  We stress that  Eq.~\eqref{eq:operationalequivalence} will be required, not Eq.~\eqref{eq:operatoridentity}. Crucially \eqref{eq:operationalequivalence} \emph{does not assume the dynamics $T_t$ is reversible}. This means our results are applicable beyond exactly unitary dynamics. For example, if $T_t = (1-s) \mathcal{U}_t + s \mathcal{D}$, with $\mathcal{D}$ depolarising noise ($\mathcal{D}(\rho) = \iden/d$ for all $\rho$), $s \in [0,1]$ and $\mathcal{U}_t$ satisfying Eq.~\eqref{eq:operatoridentity}, then Eq.~\eqref{eq:operationalequivalence} holds with $p_d = \tilde{p}_d + s(1- \tilde{p}_d)$.

We now prove the weakness encoded operationally in Eq.~\eqref{eq:operationalequivalence}, together with the observation of a $\mathcal{O}(g)$ response of a quantum system, can \emph{only} occur in the presence of contextuality.

\emph{Main theorem.} From Eq.~\eqref{eq:operationalDeltaO} and Eq.~\eqref{eq:OMprediction}, an OM predicts  that $\langle \Delta O_I \rangle_t$ equals
\small
\begin{equation}
\label{eq:OMpredictionlinearresponse}
 \! \! \! \sum_k \! o_k \! \! \left[ \int \! \!  d\lambda d\lambda' \! \mu_P(\lambda) \mathcal{T}_{T_t}(\lambda'|\lambda) \xi_{M_t}(k|\lambda') - \! \!  \int \! \! d\lambda \mu_P(\lambda) \xi_{M_t}(k|\lambda)\!\right].
\end{equation}
\normalsize
In other words, when the initial state $\ket{\psi(0)}$ is prepared a $\lambda$ is sampled with probability $\mu_P(\lambda)$; when the unitary $U_I(t)$ is performed the state is updated to $\lambda'$ with probability $ \mathcal{T}_{T_t}(\lambda'|\lambda)$; and finally a measurement of the observable $O_I(t)$ returns outcome $o_k$ with probability $\xi_{M_t}(k|\lambda')$. Then

\begin{thm}[Noncontextual bound on linear response]
Suppose the operational equivalence in Eq.~\eqref{eq:operationalequivalence} is observed. Then in any noncontextual OM
\label{thm:theorem}
\begin{equation}
\label{eq:theorembound}
|\langle \Delta O \rangle^{\rm NC}_t |\leq  4 p_d o_{\rm max},
\end{equation}
where $o_{\rm max}$ is the largest eigenvalue of $O$. 
\end{thm} 
For the proof, see SM~\ref{app:prooftheorem1}.
The only remaining idealisation in Theorem~\ref{thm:theorem} is that   Eq.~\eqref{eq:operationalequivalence} holds exactly, which will not be the case for generic noise. Luckily, this can be circumvented deploying the array of techniques developed in Ref.~\cite{mazurek16}. In summary, the experimentally realized channels may satisfy Eq.~\eqref{eq:operationalequivalence} only approximately (see SM~\ref{app:idealisation}).

 Let us now discuss the claim of the theorem. Of course in general one can have a classical linear response of $\mathcal{O}(g)$. What the main theorem proves is that a $\mathcal{O}(g)$ response, together with the phenomenology described in Eq.~\eqref{eq:operationalequivalence}, cannot be reproduced by classical models. This is because Eq.~\eqref{eq:operationalequivalence} forces noncontextual models to have a response at most of $\mathcal{O}(g^2)$.
A central question is then whether Eq.~\eqref{eq:operationalequivalence} will be observed in a quantum experiment for $g$ small enough. The next lemma gives a sufficient condition:

\begin{lem}[Operational condition test] \label{lem:sufficientcondition} Fix $t>0$ and suppose there exists $C>0$ such that the following matrix is positive definite
\begin{equation}
\tilde{J}_{kj} =  1- \frac{c_{kj}}{C}, \quad c_{kj} = (\alpha_k- \alpha_j)^2,
\end{equation}
where $\alpha_k$ are the eigenvalues of $\int_{0}^t V_I(t')dt'$. Then Eq.~\eqref{eq:operatoridentity} holds for $g$ small enough.
\end{lem}
For the proof see SM~\ref{app:operationaltest}. Note that to construct $\tilde{J}$ we only need to use linear response operators. For example, in the case of a single qubit
\[
\tilde{J}= \left[
\begin{matrix}
	1 & 1- \frac{c_{01}}{C} \\
	1- \frac{c_{01}}{C} & 1
\end{matrix}
\right]
\]
which has eigenvalues $x_1= c_{01}/C$ and $x_2 = 2- c_{01}/C$. Hence for $C$ large enough one has $\tilde{J} >0$ for any nondegenerate perturbation ($\alpha_0 \neq \alpha_1$). For a qutrit there are nontrivial counterexamples to $\tilde{J} > 0$, so one needs to perform the test for the specific scheme under consideration.

The above gives a general method to identify quantum signatures (certified against \emph{arbitrary} noncontextual  models) in arbitrary quantum systems in the linear regime:
\begin{enumerate}
\item  Compute whether $\tilde{J} > 0$. If that is the case, by carrying out the experiment one will be able to verify Eq.~\eqref{eq:operationalequivalence} (using the tools of SM~\ref{app:idealisation} to deal with noise and imperfections).
\item Check whether the response in Eq.~\eqref{eq:operationalDeltaO} is of $\mathcal{O}(g)$.
\end{enumerate}
When the two conditions above are satisfied, Theorem~\ref{thm:theorem} returns a proof of contextuality for $g$ small enough. This algorithm provides a powerful tool to identify quantum signatures. Here we apply these considerations to quantum thermodynamics and metrology.

\emph{A contextual advantage in a quantum engine.} What is the role played by nonclassicality on the performance of thermodynamic devices? Conversely, what is the ``thermodynamics of nonclassical properties'' required for the understanding of quantum devices in which thermal effects cannot be neglected? The contextuality framework offers the opportunity to rigorously investigate both questions \cite{hovhannisyan2019quantum, levy2020quasiprobability}.

Despite
recent theoretical and experimental advances, and a large number of proposals for quantum mechanical heat engines, a central outstanding question remains: Are there thermodynamic machines whose performance \emph{unavoidably} requires quantum effects? The standard comparison with a ``stochastic engine'' obtained by simple dephasing of the quantum protocol \cite{brask2015small, mitchison2015coherence, uzdin2015equivalence, levy2018quantum, camati2019coherence, dann2020quantum} is insufficient to tackle this question. The elementary example discussed in SM~\ref{app:toaster} shows that, in and by itself, the dephasing criterion is not a good notion of nonclassicality. We alternatively suggest to take contextuality as one's notion of nonclassicality and provide an upper bound on the power output of any noncontextual engine. This shows that quantum engines display a power output advantage over \emph{every} noncontextual counterpart~\footnote{The dephasing criterion claims an advantage when the quantum protocol outperforms its dephased version. Note that the dephased protocol is a special case of a noncontextual model, so our proposed criterion is more demanding than the dephasing criterion.}. 


A heat engine is a machine that works between two baths at different temperatures and whose aim is to extract work from the heat flow between the two baths. It is useful to study the functioning of an engine as a sequence of  `strokes', in which only some of the elements are involved. While the reasoning is applicable to more general models \cite{uzdin2015equivalence}, we focus here on the two-stroke engine:
\begin{enumerate}
\item The first stroke couples subsets of energy levels of the system to a hot and a cold bath to generate a non-equilibrium steady state $\rho(0)$.
\item The second stroke is a unitary driving to implement work extraction.
\end{enumerate}
We will assume that $\rho(0)$ is a two-level system, as in \cite{klatzow2019experimental}. 
Consider the work extraction process over a unitary cycle lasting an amount of time~$\tau$:
\begin{equation}
H(t) = H_0 + g V(t), \quad V(0)= V(\tau) = 0.
\end{equation}
If $U(t)$ is the unitary process generated by $H(t)$ from time $0$ to $t$, the work $W$ extracted over the cycle is
\begin{align}
W^Q  &=  \tr{}{\rho(0) H_0} - \tr{}{U(\tau) \rho(0) U^\dag(\tau) H_0} \nonumber \\
   & = \tr{}{\rho(0) H_0} - \tr{}{U_I(\tau) \rho(0) U_I^\dag(\tau) H_0}.
\end{align}
Eq.~\eqref{eq:linearresponse} returns
\begin{equation}
\label{eq:workquantum}
W^{\rm Q} = \frac{2 g \tau}{\hbar} {\rm Im} \tr{}{\rho(0) X H_0} + \mathcal{O}(g^2).
\end{equation}
where we set $X := \frac{1}{\tau} \int_{0}^\tau V_I(t) dt$ (For an interesting relation to the so-called anomalous weak values, see SM~\ref{app:awv}). Division by $\tau$ gives the power of the unitary stroke, which can be $\mathcal{O}(g)$ in the coupling strength. Furthermore, as already noted, the operational equivalence of Eq.~\eqref{eq:operationalequivalence} is satisfied generically by the unitary driving, since $\rho(0)$ is a qubit system. 
Hence, setting $E^{\rm max} = \max_i E_i$, Theorem~\ref{thm:theorem} applies. In every noncontextual model Eq.~\eqref{eq:theorembound} holds:
\begin{equation}
\label{eq:WNC}
W \leq  4 E^{\rm max} p_d :=  W_{\rm NC}.
\end{equation}
Hence $W \leq \mathcal{O}(g^2)$ as $g \rightarrow 0$ in any noncontextual model, and the same holds for power. Since we can have $W^{\rm Q}>W^{\rm NC}$ for $g$ small enough, a quantum advantage in the power output of the work stroke emerges in the weak coupling limit. In fact, the bound relies only on setting a finite upper bound on the maximum energy $E^{\rm max}$ the noncontextual model can access and not on how it represents $H_0$, $V(t)$ etc. A gap will emerge at sufficiently small $g$ (or at sufficiently short pulses for fixed~$g$).
	
	The quantum advantage is exhibited in the difference between the $\mathcal{O}(g)$ scaling possible in quantum mechanics as compared with the $\mathcal{O}(g^2)$ bound of any noncontextual model.
		This proves that the gap analysed on the basis of the dephasing criterion in Ref.~\cite{uzdin2015equivalence, garcia2019fluctuations, klatzow2019experimental, dann2020quantum} signals a true separation between classical and quantum thermodynamics. Specifically, in the presence of the phenomenology featured in the quantum experiment, the gap unavoidably requires nonclassicality in the form of  contextuality.

\emph{A contextual advantage in local metrology.} 
Local metrology is a paradigm to study the ultimate limits of parameter estimation.
We look here at the archetypal case of  phase estimation, where the relevant parameter is the phase $\eta$ in the dynamics $U_\eta = e^{-i H \eta}$ for some observable $H$. 
An initial qubit state $\ket{\psi(0)}$ is prepared, undergoes the dynamics $U_\eta$ and is measured according to some arbitrary POVM $M_x$ ($M_x \geq 0$, $\sum_x M_x = \iden$). After $N$ trials, there exists a measurement such that the error (variance) ${\rm Var}(\eta)$ in the estimated phase scales as $\mathcal{O}(1/(4 N \Delta H^2))$, where $\Delta H^2:= \bra{\psi(0)} H^2 \ket{\psi(0)} - (\bra{\psi(0)}H \ket{\psi(0)})^2$. This is finite only if the state is a (nontrivial) superposition of eigenstates of $H$, otherwise ${\rm Var}(\eta) = +\infty$. Hence dephasing trivially prevents sensing in this scheme. But what about other noncontextual models, which as already discussed can be much more complex than quantum mechanics plus dephasing? Here we show ${\rm Var}(\eta) = +\infty$ in every noncontextual model reproducing the operational phenomenology of quantum sensing. 

Let $p(x|\eta)$ be the probability of getting outcome $x$ from a measurement $M$ when the true value of the parameter is $\eta$. So $p(x|\eta) = p(x|T_\eta(P),M)$ if $P$, $T_\eta$ and $M$ are the operational descriptions of preparation, transformation and measurement procedures, represented in quantum theory by $\ket{\psi(0)}$, $U_\eta$ and $\{M_x\}$. Recall that an estimator $\hat{\eta}(x_1,x_2,\dots)$ maps the measurement outcomes $(x_1,x_2,\dots)$ to a guess  $\eta$ for the unknown parameter. For independent observations, the variance of any unbiased estimator is lower bounded by $1/(N F^{P,M}_\eta)$, with $F^{P,M}_\eta$ the Fisher information
\begin{equation}
F^{P,M}_\eta = \sum_{x} p(x|\eta) \left[\frac{\partial}{\partial \eta} \ln p(x|\eta)\right]^2.
\end{equation}
The best strategy involves optimising over all allowed preparations $P$ and measurements $M$, where for simplicity we will assume $x$ runs over a bounded, while possibly extremely large, set of indexes. In any OM, from Eq.~\eqref{eq:OMprediction}
\begin{equation}
p(x|\eta) = \int d\lambda d\lambda' \mu_P(\lambda) \mathcal{T}_{T_\eta}(\lambda'|\lambda) \xi_M(x|\lambda'),
\end{equation} 
where $\mu_P(\lambda)$, $\mathcal{T}_{T_\eta}(\lambda'|\lambda)$, $\xi_M(x|\lambda')$ are the OM descriptions of $P$, $T_\eta$ and $M$. Using the relation $p(x|\eta+\delta )= p(x| T_{\delta} (P_\eta), M)$, where $P_\eta = T_\eta(P)$, and the fact that Eq.~\eqref{eq:operationalequivalence} is satisfied with $p_d=\mathcal{O}(\delta^2)$, we can prove 
\begin{equation}
	\label{eq:fidelity}
F^{P,M}_\eta = 0
\end{equation} 
for any $P$ and any measurement $M$ with a finite number of outcomes and $p(x|\eta) \neq 0$ (see SM~\ref{app:fisher}). Hence \mbox{${\rm Var}(\eta) = +\infty$}, as anticipated. This again is a consequence of the weakness of linear response in noncontextual models. 

\emph{Outlook.} We proved that the linear response of quantum systems driven by small external perturbations has a scaling that unavoidably requires nonclassicality.
While the quantum response can scale linearly in the strength of the perturbation parameter $g$, noncontextual models reproducing the operational phenomenology in Eq.~\eqref{eq:operationalequivalence} respond only quadratically. (Curiously, one can note that since classical models display a quadratic response in the presence of the operational equivalence in Eq.~\eqref{eq:operationalequivalence}, a phenomenon such as the quantum Zeno effect is naturally expected on the basis of the assumption of noncontextuality -- see SM~\ref{app:zeno}).
 
 The $\mathcal{O}(g)$ vs $\mathcal{O}(g^2)$ gap is a certifiable quantum signature highlighting a central dynamical differences between
 noncontextual models and quantum mechanics. We gave readily applicable tools to analyze arbitrary linear response experiments. As an application, we showed that the improved performance in the power output of a quantum engine \emph{necessarily requires} nonclassical effects in the form of contextuality.

Building up on this work, it will be desirable to use the tools introduced here to reanalyse in detail the experimental heat engine signature of Ref.~\cite{klatzow2019experimental}, as well as to develop flexible certification tools applicable to larger-scale systems and based on more compelling operational constraints than `stochastic reversibility'.  While we presented an example of a performance boost unavoidably connected to contextuality, this does not settle the question of the superiority of quantum engines as practical devices. The latter requires discussing issues of scalability, control and efficiency in the implementation of basic operations. One can also envision this work as a first result in the ``thermodynamics of contextuality'', meaning how a central property signaling the departure from classical reality interacts with actual thermodynamic processes at the operational level.

{\bf Acknoledgements.} I am indebted to Matt Pusey, Ravi Kunjwal, Kavan Modi (who asked about the Zeno effect), Antonio Acin, Mohammad Mehboudi, Felix Huber, Luis Garc\'ia-Pintos, Gabriel Senno, Amikam Levy,  Joseph Bowles, David Jennings (who proposed the `Quantum Toaster Criterion'), Nicole Yunger Halpern, David Arvidsson Shukur, Mauro Paternostro and Alessio Belenchia for helpful discussions.  We acknowledge financial support from the EU Marie Sklodowska-Curie individual Fellowships (H2020-MSCA-IF-2017, GA794842), Spanish MINECO (Severo Ochoa SEV-2015-0522, project QIBEQI FIS2016-80773-P), Fundacio Cellex and Generalitat de Catalunya (CERCA Programme and SGR 875), grant EQEC No. 682726.

\onecolumngrid
\newpage
\section*{Supplemental Material of ``Certifying quantum signatures in thermodynamics and metrology via contextuality of quantum linear response''}
\subsection{Classical mechanics is noncontextual}
\label{app:hamilton}

Let's work this out explicitly in the case of preparation noncontextuality and $\lambda=(x,p)$, since the generalisation is straightforward. Suppose $P \simeq_{op} P'$, which implies $p(k|P,M) = p(k|P',M)$ for all $M$ and all outcomes $k$. By definition of OM this implies
\begin{equation*}
	\int dx dp \mu_P(x,p) \xi_M(k|x,p) = \int dx dp \mu_{P'}(x,p) \xi_M(k|x,p) \quad \forall M.
\end{equation*} 
Now, the response functions $\{ \xi_M(k|x,p)\}_M$, for varying $M$, span all indicator functions over phase space. Hence the previous equation implies $\mu_P(x,p) = \mu_{P'}(x,p)$, as required by Eq.~\eqref{eq:preparationNC} (strictly,  modulo sets of zero measure -- but likewise one should strictly define noncontextuality modulo sets of zero measure). Similarly, one can show that Eqs.~\eqref{eq:measurementNC}-\eqref{eq:transformationNC} also hold.  Hamiltonian dynamics is hence a particular member of a family of potential noncontextual emulators.

\subsection{Proof of Theorem~\ref{thm:theorem}}
\label{app:prooftheorem1}

Due to its simplicity we first give a proof under the assumption that the number of allowed $\lambda$ is finite. We then present a second proof that works in generality. The proofs follow the technique used to prove Lemma~5 in Ref.~\cite{kunjwal2019anomalous}.

\subsubsection{Proof for OM with bounded state space}

Split the sum in Eq.~\eqref{eq:OMpredictionlinearresponse} into all $\lambda,\lambda'$ with $\lambda= \lambda'$ and with $\lambda \neq \lambda'$. Using $\mathcal{T}_{T_t}(\lambda|\lambda) \leq 1$ we get

\begin{equation}
	\label{eq:oneway}
	\langle \Delta O_I \rangle_t \leq \sum_k o_k  \sum_{\lambda \neq \lambda'} \mu_P(\lambda) \mathcal{T}_{T_t}(\lambda'|\lambda) \xi_{M_t}(k|\lambda').
\end{equation}
Now use the noncontextuality assumption of Eq.~\eqref{eq:transformationNC}. Eq.~\eqref{eq:operationalequivalence} requires
\begin{equation}
	\label{eq:noncontextualityapplied}
	\frac{1}{2} \mathcal{T}_{T_t}(\lambda'|\lambda) +  \frac{1}{2} \mathcal{T}_{T^*_t}(\lambda'|\lambda) = (1-p_d) \delta_{\lambda \lambda'} + p_d \mathcal{T}_{T'_t}(\lambda'|\lambda).
\end{equation}
Note that the transition probability associated to $T_{\rm id}$ in noncontextual models is a Kronecker delta, since one way to realize $T_{\rm id}$ is to let no time pass. Hence, for $\lambda \neq \lambda'$, $\mathcal{T}_{T_t}(\lambda'|\lambda) \leq 2 p_d \mathcal{T}_{T'_t}(\lambda'|\lambda)$, which shows that the transition probabilities between distinct $\lambda$s must be of $O(p_d)$. Then,
\begin{equation*}
	\langle \Delta O_I \rangle_t \leq 2p_d \sum_k o_k  \sum_{\lambda \neq \lambda'} \mu_P(\lambda) \mathcal{T}_{T'_t}(\lambda'|\lambda) \xi_{M_t}(k|\lambda') \leq 2p_d o_{\rm max},
\end{equation*}
where in the last inequality we extended the sum to all $\lambda$, $\lambda'$ and used that 
\begin{equation}
	\sum_{\lambda, \lambda'} \mu_P(\lambda) \mathcal{T}_{T'_t}(\lambda'|\lambda) \xi_{M_t}(k|\lambda') = p(k|T'_t(P),M_t), \quad \sum_k p(k|T'_t(P),M_t) =1.
\end{equation}

For the converse, let us start with Eq.~\eqref{eq:OMpredictionlinearresponse} for $\langle \Delta O_I \rangle_t$ and use Eq.~\eqref{eq:noncontextualityapplied} to replace $\mathcal{T}_{T_t}(\lambda'|\lambda)$:
\small
\begin{align*}
	\! \! \! \! \! \! - \langle \Delta O_I \rangle_t \! = & \! \sum_k o_k \! \! \left[\!  \sum_\lambda \mu_P(\lambda) \xi_{M_t}(k|\lambda) \! + \! \! \sum_{\lambda,\lambda'} \mu_P(\lambda) \mathcal{T}_{T^*_t}(\lambda'|\lambda) \xi_{M_t}(k|\lambda') - 2(1-p_d) \! \sum_\lambda  \mu_P(\lambda) \xi_{M_t}(k|\lambda) - 2 p_d \! \sum_{\lambda,\lambda'} \mu_P(\lambda) \mathcal{T}_{T'_t}(\lambda'|\lambda) \xi_{M_t}(k|\lambda') \! \right] \\
	\leq & \sum_k o_k \left[ \sum_{\lambda,\lambda'} \mu_P(\lambda) \mathcal{T}_{T^*_t}(\lambda'|\lambda) \xi_{M_t}(k|\lambda') - \sum_\lambda \mu_P(\lambda) \xi_{M_t}(k|\lambda)  + 2p_d \sum_\lambda  \mu_P(\lambda) \xi_{M_t}(k|\lambda)\right],
\end{align*}
\normalsize
where in the second line we dropped the term $ - 2 p_d \sum_{\lambda,\lambda'} \mu_P(\lambda) \mathcal{T}_{T'_t}(\lambda'|\lambda) \xi_{M_t}(k|\lambda')$. We have
\begin{equation*}
	2p_d \sum_k o_k  \sum_\lambda  \mu_P(\lambda) \xi_{M_t}(k|\lambda) = 2p_d \sum_k o_k p(k|P,M_t) \leq 2p_d o_{\rm max},
\end{equation*}
so
\small
\begin{align*}
	- \langle \Delta O_I \rangle_t \leq & \sum_k o_k \left[ \sum_{\lambda,\lambda'} \mu_P(\lambda) \mathcal{T}_{T^*_t}(\lambda'|\lambda) \xi_{M_t}(k|\lambda') - \sum_\lambda \mu_P(\lambda) \xi_{M_t}(k|\lambda)\right]  + 2p_d o_{\rm max}
\end{align*}
\normalsize
Splitting the sum in the first term into $\lambda,\lambda'$ with $\lambda= \lambda'$ and with $\lambda \neq \lambda'$ and using $\mathcal{T}_{T^*_t}(\lambda|\lambda) \leq 1$ we get an equation almost identical to Eq.~\eqref{eq:oneway}:
\begin{equation}
	- \langle \Delta O_I \rangle_t \leq \sum_k o_k  \sum_{\lambda \neq \lambda'} \mu_P(\lambda) \mathcal{T}_{T^*_t}(\lambda'|\lambda) \xi_{M_t}(k|\lambda') + 2p_d o_{\rm max}.
\end{equation}
Due to the symmetric role played by $T_t$ and $T^*_t$, we can repeat the same proof we used to bound $\langle \Delta O_I \rangle_t$ to get a bound on the first term in the previous  equation as
\begin{equation*}
	\sum_k o_k  \sum_{\lambda \neq \lambda'} \mu_P(\lambda) \mathcal{T}_{T^*_t}(\lambda'|\lambda) \xi_{M_t}(k|\lambda') \leq 2p_d o_{\rm max}.
\end{equation*}
This allow us to conclude $- \langle \Delta O_I \rangle_t \leq 4 p_d o_{\rm max}$.

\subsubsection{Proof OM with unbounded state space}

Combining Eqs.~\eqref{eq:OMprediction} and \eqref{eq:operationalDeltaO} in the case of a potentially continuous number of states in the OM we get
\begin{equation}
	\label{eq:firsteqappendix}
	\langle \Delta O_I \rangle_t = \sum_k o_k \left[\int d\lambda \int d\lambda' \mu_P(\lambda) \mathcal{T}_{T_t}(\lambda'|\lambda) \xi_{M_t}(k|\lambda') - \int d\lambda  \mu_P(\lambda) \xi_{M_t}(k|\lambda)\right].
\end{equation}
For each outcome $k$ and $\lambda$ we can partition the space of ontological states $\Lambda$ as
\begin{equation}
	\Lambda^k_\lambda =\{\lambda'|\xi_{M_t} (k|\lambda') > \xi_{M_t}(k|\lambda)\}, \quad \bar{\Lambda}^k_\lambda = \Lambda \backslash \Lambda^k_\lambda.
\end{equation} 
Then
\small
\begin{align}
	\langle \Delta O_I \rangle_t & =  \sum_k o_k \left[\int d\lambda \int_{\Lambda^k_\lambda} d\lambda' \mu_P(\lambda) \mathcal{T}_{T_t}(\lambda'|\lambda) \xi_{M_t}(k|\lambda') + \int d\lambda \int_{\bar{\Lambda}^k_\lambda} d\lambda' \mu_P(\lambda) \mathcal{T}_{T_t}(\lambda'|\lambda) \xi_{M_t}(k|\lambda') - \int d\lambda  \mu_P(\lambda) \xi_{M_t}(k|\lambda)\right] \nonumber \\
	& \leq \sum_k o_k \left[\int d\lambda \int_{\Lambda^k_\lambda} d\lambda' \mu_P(\lambda) \mathcal{T}_{T_t}(\lambda'|\lambda) \xi_{M_t}(k|\lambda') + \int d\lambda \int_{\bar{\Lambda}^k_\lambda} d\lambda' \mu_P(\lambda) \mathcal{T}_{T_t}(\lambda'|\lambda) \xi_{M_t}(k|\lambda) - \int d\lambda  \mu_P(\lambda) \xi_{M_t}(k|\lambda)\right] \nonumber \\
	& = \sum_k o_k \left[\int d\lambda \int_{\Lambda^k_\lambda} d\lambda' \mu_P(\lambda) \mathcal{T}_{T_t}(\lambda'|\lambda) [\xi_{M_t}(k|\lambda') -\xi_{M_t}(k|\lambda)] + \int d\lambda \int d\lambda' \mu_P(\lambda) \mathcal{T}_{T_t}(\lambda'|\lambda) \xi_{M_t}(k|\lambda) - \int d\lambda  \mu_P(\lambda) \xi_{M_t}(k|\lambda)\right] \nonumber \\
	& = \sum_k o_k \int d\lambda \int_{\Lambda^k_\lambda} d\lambda' \mu_P(\lambda) \mathcal{T}_{T_t}(\lambda'|\lambda) [\xi_{M_t}(k|\lambda') -\xi_{M_t}(k|\lambda)]  \nonumber \\
	& \leq \sum_k o_k \int d\lambda \int d\lambda' \mu_P(\lambda) [\mathcal{T}_{T_t}(\lambda'|\lambda) + \mathcal{T}_{T^*_t}(\lambda'|\lambda)][\xi_{M_t}(k|\lambda') -\xi_{M_t}(k|\lambda)]  \nonumber \\
	& = 2p_d \sum_k o_k \int d\lambda \int d\lambda' \mu_P(\lambda) \mathcal{T}_{T'_t}(\lambda'|\lambda) [\xi_{M_t}(k|\lambda') -\xi_{M_t}(k|\lambda)]  \nonumber 
	\\
	& \leq 2p_d \sum_k o_k \int d\lambda \int d\lambda' \mu_P(\lambda) \mathcal{T}_{T'_t}(\lambda'|\lambda) \xi_{M_t}(k|\lambda') \leq 2p_d o_{\rm max},
	\label{app:proofsteps}
\end{align}
\normalsize
where we used: $\xi_{M_t}(k|\lambda') \leq \xi_{M_t}(k|\lambda)$ in $\bar{\Lambda}^k_\lambda$ to get the second line; $\int d\lambda' T_{T_t}(\lambda'|\lambda)=1$ to get the fourth line; and to get the sixth line we used noncontextuality applied to the operational equivalence of Eq.~\eqref{eq:operationalequivalence} in the main text, i.e.
\begin{equation}
	\label{eq:noncontextualityappliedcontinuous}
	\mathcal{T}_{T_t}(\lambda'|\lambda) + \mathcal{T}_{T^*_t}(\lambda'|\lambda) = 2(1-p_d) \delta(\lambda-\lambda') + 2p_d T_{T'_t}(\lambda'|\lambda). 
\end{equation}
For the last inequality we used
\begin{equation*}
	\int d\lambda \int d\lambda' \mu_P(\lambda) \mathcal{T}_{T'_t}(\lambda'|\lambda) \xi_{M_t}(k|\lambda') = p(k|T'_t(P),M_t), \quad \sum_k p(k|T'_t(P),M_t) =1.
\end{equation*}

The proof of  $- \langle \Delta O_I \rangle_t \leq 2 p_d o_{\rm max}$ involves the same idea we used for the case of a finite state space in the previous subsection. We start by using Eq.~\eqref{eq:noncontextualityappliedcontinuous} to replace $\mathcal{T}_{T_t}(\lambda'|\lambda)$ in Eq.~\eqref{eq:firsteqappendix}. Doing so and neglecting a negative term we get
\begin{equation*}
	- \langle \Delta O_I \rangle_t = \sum_k o_k \left[\int d\lambda \int d\lambda' \mu_P(\lambda) \mathcal{T}_{T^*_t}(\lambda'|\lambda) \xi_{M_t}(k|\lambda') - \int d\lambda  \mu_P(\lambda) \xi_{M_t}(k|\lambda)\right] + 2p_d \sum_k o_k \int d\lambda \mu_P(\lambda) \xi_{M_t}(k|\lambda). 
\end{equation*}
As before, $2p_d \sum_k o_k \int d\lambda \mu_P(\lambda) \xi_{M_t}(k|\lambda) \leq 2p_d o_{\rm max}$. Furthermore, noting the symmetric role played by $T_t$ and $T^*_t$ in Eq.~\eqref{eq:noncontextualityappliedcontinuous}, we can reapply the first part of the proof to obtain the bound
\begin{equation*}
	\sum_k o_k \left[\int d\lambda \int d\lambda' \mu_P(\lambda) \mathcal{T}_{T^*_t}(\lambda'|\lambda) \xi_{M_t}(k|\lambda') - \int d\lambda  \mu_P(\lambda) \xi_{M_t}(k|\lambda)\right] \leq 2p_d o_{\rm max}.
\end{equation*}
We conclude $	- \langle \Delta O_I \rangle_t \leq 4 p_d o_{\rm max}$.

\subsection{Proof of Lemma~\ref{lem:sufficientcondition}}
\label{app:operationaltest}

Fix $t>0$ and let $\mathcal{U}_g = U_g (\cdot) U_g^\dag$ with $U_g = \mathbb{I} - \frac{ig}{\hbar}\int_{0}^t V_I(t')dt' + O(g^2)$. We have $U_g = e^{- i H_g}$ for some Hermitian operator $H_g$. By Hadamard lemma $\mathcal{U}_g = e^{-i \mathcal{H}_g}$, with $\mathcal{H}_g(X) = [H_g,X]$. Then
\begin{equation}
	\frac{1}{2}\mathcal{U}_g + \frac{1}{2}\mathcal{U}^\dag_g = \frac{1}{2}e^{-i \mathcal{H}_g} + \frac{1}{2}e^{i \mathcal{H}_g} = \mathcal{I} + \sum_{n=1}^\infty \frac{i^{2n}}{(2n)!}\mathcal{H}^{2n}_g = (1-p_d)\mathcal{I} + p_d\left(\mathcal{I} + \frac{1}{p_d}\sum_{n=1}^\infty \frac{(-1)^n}{(2n)!}\mathcal{H}^{2n}_g\right).
\end{equation}    
Set $\mathcal{C}_g = \mathcal{I} + \frac{1}{p_d} \sum_{n=1}^\infty \frac{(-1)^n}{(2n)!}\mathcal{H}^{2n}_g$. We want to prove that for $g$ small enough $\mathcal{C}_g$ is a quantum channel, i.e. a Completely Positive Trace Preserving (CPTP) map, for some choice of $p_d = O(g^2)$. 
\begin{itemize}
	\item Trace-preserving condition:
\end{itemize}
\begin{equation}
	{\rm Tr}[\mathcal{C}_g(\rho)] = 1 + \frac{1}{p_d} \sum_{n=1}^\infty \frac{(-1)^n}{(2n)!}{\rm Tr}[\mathcal{H}^{2n}_g(\rho)]. 
\end{equation}
However
\begin{align}
	{\rm Tr}[\mathcal{H}_g(\rho)] =  {\rm Tr}[H_g \rho - \rho H_g] = 0, \quad \quad
	{\rm Tr}[\mathcal{H}^k_g(\rho)] =  {\rm Tr}[ \mathcal{H}^{k-1}_g(H_g \rho - \rho H_g)],
\end{align}
so proceeding by induction one has ${\rm Tr}[\mathcal{H}^n(\rho)] = 0$ for all $n$. Hence ${\rm Tr}[\mathcal{C}_g(\rho)] = 1$, so $\mathcal{C}_g$ is trace-preserving.
\begin{itemize}
	\item Complete Positivity:
\end{itemize}
Let $\ket{\phi^+} = \frac{1}{\sqrt{d}} \sum_k \ket{kk}$, with $H_g \ket{k} = E_k(g) \ket{k}$ (with an implicit dependence of $\ket{k}$ on $g$). By the Choi-Jamio\l kowski isomorphism the condition of complete positivity is equivalent to $J_g := \mathcal{C}_g \otimes \mathcal{I} (\ketbra{\phi^+}{\phi^+})/d \geq 0$ for $g$ small enough.
\begin{align}
	J_g & = \frac{1}{d} \sum_{k,j} \mathcal{C}_g(\ketbra{k}{j}) \otimes \ketbra{k}{j} = \frac{1}{d}\sum_{k,j} \left[ \ketbra{k}{j} \otimes \ketbra{k}{j} + \frac{1}{p_d} \sum_{n=1}^\infty \frac{(-1)^n}{(2n)!}\mathcal{H}^{2n}(\ketbra{k}{j}) \otimes \ketbra{k}{j}  \right] \\
	& =\frac{1}{d}\sum_{k,j} \left[ \ketbra{k}{j} \otimes \ketbra{k}{j} + \frac{1}{p_d} \sum_{n=1}^\infty \frac{(-1)^n}{(2n)!}(E_k(g) - E_j(g))^{2n}(\ketbra{k}{j}) \otimes \ketbra{k}{j}  \right],
\end{align}
where in the last equation we used $\mathcal{H}_g(\ketbra{k}{j}) = H_g \ketbra{k}{j} - \ketbra{k}{j}H_g = (E_k(g) - E_j(g))\ketbra{k}{j}$. Now, using \mbox{$\cos x = \sum_{n=0}^\infty x^{2n}/(2n)!$},
\begin{align}
	J_g = \frac{1}{d}\sum_{k,j} \left[ \left(1-\frac{1}{p_d}\right) + \frac{1}{p_d} \cos(E_k(g)-E_j(g)) \right] \ketbra{k}{j} \otimes \ketbra{k}{j}. 
\end{align}
Note that $H_g = i \log[ \mathbb{I} - i \frac{g}{\hbar} \int_{0}^t V_I(t')dt'+ O(g^2)] = \frac{g}{\hbar} \int_{0}^t V_I(t')dt' + O(g^2)$. Hence, if we denote by $\alpha_i$ the eigenvalues of $\int_{0}^t V_I(t')dt'$ (which do not depend on $g$), the eigenvalues of $H_g$ will be $\frac{g}{\hbar} \alpha_k + O(g^2)$. This follows from the fact that the eigenvalues of $H_g$ can only differ from those of $\frac{g}{\hbar}\int_{0}^t V_I(t') dt'$ by $O(g^2)$ \cite{kahan1975spectra, ipsen2009refined}.

Hence \mbox{$E_k(g) - E_j(g) = \frac{g}{\hbar} (\alpha_k - \alpha_j) + O(g^2)$}.
We have
\begin{equation}
	\cos[E_k(g) - E_j(g)] = \cos \left[\frac{g}{\hbar} (\alpha_k - \alpha_j) + O(g^2)\right] = 1 - \frac{g^2}{2\hbar^2}(\alpha_k - \alpha_j)^2 + O(g^4). 
\end{equation}
This gives
\begin{equation}
	J_g = \frac{1}{d}\sum_{k,j} \left[ 1- \frac{g^2 (\alpha_k - \alpha_j)^2}{2 \hbar^2 p_d} + \frac{O(g^4)}{p_d} \right] \ketbra{kk}{jj}.
\end{equation}
Set $p_d = C g^2/ (2 \hbar^2) $ for some constant $C>0$. Hence, denoting $c_{kj} = (\alpha_k - \alpha_j)^2$,
\begin{equation}
	J_g = \frac{1}{d}\sum_{k,j} \left[ 1- \frac{c_{kj}}{C} + O(g^2) \right] \ketbra{kk}{jj} = \frac{1}{d}\tilde{J} + O(g^2).
\end{equation}
We study the matrix
\begin{equation}
	\tilde{J} = \sum_{k,j} \left[ 1- \frac{c_{kj}}{C}\right] \ketbra{kk}{jj}.
\end{equation}
The eigenvalues of $J_g$ can only differ from those of $\tilde{J}$ by $O(g^2)$ \cite{kahan1975spectra, ipsen2009refined}. Hence, if we can find a constant $C>0$ such that $\tilde{J}>0$, then $J_g > 0$ for $g$ small enough.

\subsection{Elementary counterexamples to the validity of the dephasing criterion: the ``quantum toaster''}
\label{app:toaster}

A widely used criterion to witness a quantum signature in thermodynamics is to compare a protocol with its ``dephased version''. That is, one runs a given protocol first on a quantum system displaying quantum coherence, then on the same system in which coherence is removed. If we see a lower performance in the latter case as compared to the former, the criterion goes, we have a quantum signature or even a quantum advantage. 

However, it is not difficult to build elementary counterexamples to the general validity of this criterion. Other authors presented counterexamples in nontrivial settings \cite{nimmrichter2017quantum, onam2019classical}, but it is perhaps useful to analyse an almost trivial case that highlights the problem starkly.

Consider a classical engine (maybe a toaster) whose work stroke occurs only if a single photon hits a switch that turns the engine on for  a single stroke. Suppose single photons are generated by a source at a fixed rate $r=1/\tau$ (with $\tau$ larger than the duration of an engine stroke) and sent through a standard Mach-Zender interferometer. We call the bright output port of the interferometer ``port 1'' and attach it to the switch that turns on the classical engine for a single stroke. Each single photon goes through the interferometer in a superposition of the two branches and, by quantum interference, is found at the output port $1$ with probability $1$. Hence, each photon switches the classical engine on for a stroke. Under this protocol, in a unit time the engine undergoes $r$ strokes, so the (average) power is proportional to $r$. 

Now compare this protocol to its decohered version, obtained by simply destroying quantum superposition in the interferometer by means of a non-destructive measurement of the which-way information. The photon is now found with probability $1/2$ in the upper branch and with probability $1/2$ in the lower branch. Hence, the photon comes out $1/2$ of the times from port $1$ (activating the engine) and $1/2$ of the times from port $2$ (doing nothing). This implies that, after destroying quantum superposition, the power output of the engine is halved. A naive application of the dephasing criterion identifies this as a quantum feature of the engine related to a power advantage. However, clearly the quantum effect is not instrumental to the improved performance, since there is a trivial classical emulation of the engine that uses no superposition: just remove the interferometer and send each photon directly to the engine switch. 

Typically the dephasing criterion is applied to highly nontrivial machines for which we do not know if a classical emulation exists or not. Nevertheless what the discussion above highlights  is that, in and by itself, the dephasing criterion is insufficient to prove the absence of  a classical model displaying the exact same signature. Hence the necessity of rigorous certificates, the main topic of this work.

\subsection{Dealing with idealisations}
\label{app:idealisation}

The main theorem assumes the exact validity of Eq.~\eqref{eq:operationalequivalence}. Luckily, as demonstrated in Ref.~\cite{mazurek16}, it is not necessary that the experimentally achieved channels satisfy the corresponding equality exactly. While the techniques to remove this idealisation are discussed extensively in~\cite{mazurek16}, here we give a summary.

In this context, the experimentally realisable dynamics are called \emph{primary}. These can be characterised by assuming access to a tomographically complete set of measurements and preparations. Then one looks in the convex hull of these primary channels for new channels (called \emph{secondary}) that satisfy Eq.~\eqref{eq:operationalequivalence} exactly. Note in passing that, depending on the specifics of the experiment, one may have to introduce an auxiliary primary channel to make sure that the convex hull of the primary channels contains the required secondary channels. Then one applies Theorem~1 to the secondary channels to obtain a proof of contextuality. The price one pays is that this procedure will lead to larger values for $p_d$ and hence to weaker bounds in Theorem~\ref{thm:theorem}. Hence, as expected the noise needs to be suitably small compared to the actual dynamics of interest. In conclusion, given a concrete experimental setup, the techniques of Ref.~\cite{mazurek16} provide clear strategies to deal with idealisations.

\subsection{Relation to weak values and quasiprobabilities.} 
\label{app:awv}

We proved that the quantum mechanical operation of the two-stroke engine cannot be emulated by classical means. Here we gather some more intuition as to why this is the case. It is useful to recast Eq.~\eqref{eq:workquantum} in two alternative ways. On the one hand, setting $H_I(t) := e^{i H_0 t/\hbar}(H_0 + g V(t))e^{-i H_0 t/\hbar} = H_0 + g V_I(t)$ one has
\begin{equation}
	W^{\rm Q} = \frac{g}{i \hbar} \int_{0}^\tau \tr{}{[H_I(0),H_I(t)] \rho(0)}, 
\end{equation}  
i.e. $W^{\rm Q}$ in the weak coupling regime is related to two-points correlations functions, as expected from linear response theory. On the other hand such correlation function can also be rewritten as ($X := \frac{1}{\tau} \int_{0}^\tau V_I(t) dt$, $H_0 = \sum_i E_i \Pi_{E_i}$)
\begin{equation}
	\label{eq:workweakvalues}
	W^Q = \frac{2 g \tau}{\hbar} \sum_i \tr{}{\Pi_{E_i} \rho(0)}  E_i \,  {\rm Im}\left[{}_{E_i}\langle X \rangle_{\rho(0)} \right] + O(g^2).
\end{equation}
where ${}_{E_i}\langle X \rangle_{\rho(0)} :=  \frac{\tr{}{\Pi_{E_i} X \rho(0)}}{\tr{}{\Pi_{E_i} \rho(0)} }$, i.e. a weighted sum of imaginary parts of so-called \emph{generalized weak values} \cite{wiseman2002weak} (the quantities ${}_{E_i}\langle X \rangle_{\rho(0)} $). It was previously noticed that the latter characterise the first order response of the probability of obtaining outcome $E_i$ due to a unitary driving \cite{dressel2014colloquium}. As we can see, that is due to their relation to correlation functions. Furthermore, Ref.~\cite{kunjwal2019anomalous} has shown that, in finite-dimensional quantum systems, the weak measurement protocols that are used to estimate the quantities ${\rm Im}{}_{E_i}\langle X \rangle_{\rho(0)}$ do not admit a classical, i.e. noncontextual emulation, due to their specific information-disturbance tradeoff. One is then lead to conjecture that the work and power in Eq.~\eqref{eq:workweakvalues} may themselves be -- in some sense to be made precise -- nonclassical. We have proved that this is indeed the case. Alternatively, note that 
\begin{equation}
	W^Q = \frac{2 g \tau}{\hbar} \sum_i E_i \,  {\rm Im}\left[\tr{}{\Pi_{E_i} X \rho(0)} \right] + O(g^2),
\end{equation}
i.e. the first order response of $W$ is directly related to the imaginary part of the Kirkwood-Dirac quasiprobability.

\subsection{Proof of Eq.~\eqref{eq:fidelity}: $F^{P,M}_\eta = 0$.} 
\label{app:fisher}
By definition,
\begin{equation}
	F_{\eta} = \lim_{\delta \rightarrow 0} \sum_{x} p(x|\eta) \left[ \frac{\ln p(x|\eta + \delta) - \ln p(x|\eta)}{\delta} \right]^2,
\end{equation}
\begin{equation}
	p(x|\eta) = p(x|P_\eta,M) = \int d\lambda \mu_{P_\eta}(\lambda) \xi_M(x|\lambda), \quad p(x|\eta+\delta) = p(x|T_\delta(P_\eta), M) = \int d\lambda d\lambda' \mu_{P_\eta}(\lambda)\mathcal{T}_{T_\delta}(\lambda'|\lambda) \xi_M(x|\lambda').
\end{equation}
It follows that the difference $p(x|\eta+\delta) - p(x|\eta)$ has the same form as Eq.~\eqref{eq:firsteqappendix} with a single $o_k =1$ and all others equal to zero. We can connect to the unitary driving scenario of Eq.~\eqref{eq:unitarydriving} by setting $H_0 = 0$, $t= \eta$ and $V(\eta) = H$. Since $V_I(\eta) = H$ and we are considering a qubit system, for any $H$ Lemma~\ref{lem:sufficientcondition} ensures that the operational equivalence in Eq.~\eqref{eq:operationalequivalence} is satisfied with $T_\eta$ described in quantum mechanics by the unitary $U_\eta = e^{-i H \eta}$ and $T^*_\eta$ by $U^\dag_\eta$. 

Hence the derivation in Appendix~\ref{app:prooftheorem1} gives
\begin{equation}
	|p(x|\eta + \delta) - p(x|\eta)| \leq 2p_d = \mathcal{O}(\delta^2).
\end{equation}

Since by assumption $p(x|\eta) \neq0$, we consider the expression $\ln^2 [p(x|\eta + \delta)/ p(x|\eta)]$. $\ln^2 y$ is monotonically decreasing for $y \leq 1$ and increasing for $y \geq 1$, so we distinguish two cases. 
\begin{itemize}
	\item If $p(x|\eta + \delta) \geq p(x|\eta)$,
	\begin{align*}
		\ln^2 \frac{p(x|\eta + \delta)}{p(x|\eta)} \leq \ln^2 \frac{|p(x|\eta + \delta) - p(x|\eta)| + p(x|\eta)}{ p(x|\eta)} \leq \ln^2 \left[1 + \frac{2 p_d}{ p(x|\eta)}\right]
	\end{align*} 
	
	\item If $p(x|\eta + \delta) < p(x|\eta)$ and for $\delta$ small enough,
	\begin{align*}
		\ln^2 \frac{p(x|\eta + \delta)}{p(x|\eta)} \leq \ln^2 \frac{-|p(x|\eta + \delta) - p(x|\eta)| + p(x|\eta)}{ p(x|\eta)} \leq \ln^2 \left[1 - \frac{2 p_d}{ p(x|\eta)}\right]
	\end{align*} 
\end{itemize}
Since  $\ln^2[1-y] \geq \ln^2[1+y]$ for $y \in [0,1]$, we conclude that in both cases, assuming $p_d$ is small enough,
\begin{equation}
	\frac{p(x|\eta)}{\delta^2} \ln^2 \frac{p(x|\eta + \delta)}{p(x|\eta)} \leq \frac{p(x|\eta)}{\delta^2} \ln^2 \left[1 - \frac{2 p_d}{ p(x|\eta)}\right] = \frac{4 p_d^2}{\delta^2 p(x|\eta)} + \frac{O(p_d^3)}{\delta^2}.
\end{equation}
Using $p_d=\mathcal{O}(\delta^2)$,
\begin{equation}
	F_{\eta} = \lim_{\delta \rightarrow 0} \sideset{}{'}\sum_{x} \frac{4 }{p(x|\eta)} \times \mathcal{O}(\delta^2) = 0
\end{equation}
Note that we assumed the number of outcomes to be finite.

\subsection{Zeno effect from noncontextuality} 
\label{app:zeno}

One can also reverse the discussion and analyse phenomena which, while potentially surprising, are nonetheless naturally expected in noncontextual models. This is the case of the quantum Zeno effect. Consider a qubit system undergoing Rabi oscillations under a unitary $U_t$. It is well-known that, if the system is prepared in the initial state $\ket{0}$ and continuously measured in the computational basis, it has vanishing probability of being found in the state $\ket{1}$, so that monitoring ``freezes'' the evolution. 

In a generic OM the experiment is described as follows. A $\lambda$ is sampled according to some probability distribution $\mu_{P_0}(\lambda)$, where $P_0$ is the preparation described in quantum mechanics by $\ket{0}$. Then the state is updated according to matrix of transition probabilities $\mathcal{T}_{T_t}(\lambda'|\lambda)$, where $T_t$ is the transformation described in quantum mechanics by $U_t$. Finally the system is measured according to the procedure $M$ (in quantum mechanics, the projective measurement in the $\{\ket{0}, \ket{1}\}$ basis). In the OM the measurement is associated to a response function $\xi_{M}(k|\lambda)$, which gives the probability outcome $k=0,1$ is returned if the state is $\lambda$. We are interested in what happens when we measure many  times $M$ in a given time interval.

From our analysis a generic feature of noncontextual OM respecting the operational feature in Eq.~\eqref{eq:operationalequivalence} is that the stochastic processes associated to quantum unitary dynamics induce a change $\lambda \mapsto \lambda'$ ($\lambda \neq \lambda'$) with a probability which is quadratic (rather than linear) in the interaction time $t$ when $t$ is small (see proof of Theorem~\ref{thm:theorem} in Appendix~\ref{app:prooftheorem1}). 

Given that, consider the following simple noncontextual model of the quantum Zeno experiment. Take
\begin{equation}
	\mathcal{T}_{T_t}(\lambda'|\lambda) = (1-p_d(t)) \delta_{\lambda \lambda'} + p_d(t) \mathcal{T}_{T'_t}(\lambda'|\lambda)
\end{equation}  for some $\mathcal{T}_{T'_t}$. Similarly take the matrix of transition probabilities associated to the transformation $T^*_t$ (in quantum mechanics $U^\dag_t$) to be
\begin{equation}
	\mathcal{T}_{T^*_t}(\lambda'|\lambda) = (1-p_d(t)) \delta_{\lambda \lambda'} + p_d(t) \mathcal{T}_{T^{''}_t}(\lambda'|\lambda)
\end{equation}
for some $\mathcal{T}_{T^{''}_t}$ . Since Eq.~\eqref{eq:operationalequivalence} is fulfilled, the assumption of noncontextuality requires $p_d(t) = O(t^2)$ when $t \rightarrow 0$. Given this, let's see how one can naturally completes the model so that it displays a Zeno effect.

Let $\mu_{P_x} (\lambda)$  $x=0,1$ be the (disjoint) probability distributions associated to $\ket{0}$, $\ket{1}$. We will take $\xi_M(x|\lambda) =1$ on the support of $\mu_{P_x}$ and zero otherwise. Furthermore, the post-measurement state is simply given by $\mu_{P_x}(\lambda)$ if the outcome is $x$. Since the state is initialized in $\ket{0}$, the survival probability $p(x=0|T_t(P_0), M)$ is bounded by 
\begin{equation}
	p(x=0|T_t(P_0), M) = \sum_{\lambda,\lambda'}\mu_{P_0}(\lambda)\mathcal{T}_{T_t}(\lambda'|\lambda)\xi_M(0|\lambda') \geq 1-p_d(t) = 1- c t^2
\end{equation}
where we took $p_d(t) = c t^2$ for some constant $c \geq 0$. The post-measurement state is $\mu_{P_0}$ if the measurement returns $x=0$, and $\mu_{P_1}$ in case of outcome $x=1$. If we divide the total evolution time $\tau$ into $N$ steps and perform a measurement after each, the overall survival probability will scale as $(1-c \tau^2/N^2)^N=1-c \tau^2/N + O(1/N^2)$, which goes to $1$ as the number of measurements $N \rightarrow \infty$. The underlying reason is the quadratic scaling of $\mathcal{T}_{T_t}(\lambda'|\lambda)$ for $\lambda'\neq \lambda$, itself a consequence of noncontextuality. A Zeno effect emerges from the assumption of noncontextuality of the underlying OM. 

In summary, we have presented a natural noncontextual OM displaying a Zeno effect. The intuition behind is that -- in noncontextual OM -- unitaries close to the identity in quantum mechanics must be represented by stochastic processes that induce a change of state $\lambda \mapsto \lambda'$ with probability quadratic in the interaction time. In such models a projective measurement will just reprepare the initial state with a probability that departs from $1$ only by quadratic corrections. By ideally performing an increasingly large number of such measurements within a fixed time interval, the overall probability of inducing a change in the initial state becomes vanishingly small.

This is certainly compatible with the fact that the Zeno effect is related to measurement disturbance, a feature that is possible to reproduce in noncontextual models \cite{spekkens2007evidence}. At the same time, the Zeno effect is often considered a purely nonclassical phenomenon. In this respect our contribution does not rule out that purely quantum features in the Zeno effect exist. However the existence of a noncontextual model reproducing the core  operational features of the Zeno effect forces us to be precise as to what exactly is nonclassical about it. For an idea of how this problem may be approached, see for example Ref.~\cite{schmid2018contextual} (Ref.~\cite{lostaglio2019contextual}), where nonclassical effects are identified in the context of state discrimination (quantum cloning) despite the existence of models \cite{spekkens2007evidence} reproducing some of its core features. Similarly, there may be specific aspects of the Zeno effect that cannot be explained in noncontextual models. We leave this question open for future research.

\bibliography{Bibliography}
\bibliographystyle{apsrev4-2}

\end{document}